\documentclass[twoside,12pt]{article}
\usepackage{epsf,epsfig,amssymb,amsmath,graphicx,float}
\usepackage{color}
\usepackage{hyperref}
\usepackage{indentfirst}
\hypersetup{colorlinks,citecolor=blue,linkcolor=red,urlcolor=red}

\usepackage{amsmath}

\newcommand{\lsim}{\raisebox{-0.13cm}{~\shortstack{$<$ \\[-0.07cm] $\sim$}}~}
\newcommand{\gsim}{\raisebox{-0.13cm}{~\shortstack{$>$ \\[-0.07cm] $\sim$}}~}

\begin{document}
\renewcommand{\thefootnote}{\fnsymbol{footnote}}

\begin{titlepage}

\begin{center}

\vspace{1cm}

{\Large {\bf Constraints on Asymmetric Dark Matter Self Annihilation Cross
    Sections in Non-standard Cosmological Scenarios  }}

\vspace{1cm}

{\bf Fangyu Liu, Hoernisa Iminniyaz\footnote{Corresponding 
author, wrns@xju.edu.cn}}

\vskip 0.15in
{\it
{School of Physics Science and Technology, Xinjiang University, \\
Urumqi 830017, China} \\
}

\abstract{ We investigate the relic abundance of asymmetric dark matter in the
  non-standard cosmological scenarios when the annihilation cross section
  includes self annihilations. Here we discuss the kination model and brane
  world cosmology. When the self annihilation is permitted for
  asymmetric dark matter, there is possibility of washing out the pre-existed
  asymmetry. We find the constraints on the cross section to avoid the
  complete washing out of the asymmetry in the non-standard cosmological
  scenarios. The enhanced cosmic expansion rate causes the freeze out point of
  wash-out to be
  earlier. The larger self annihilation cross sections are allowed to exist in
  kination model and brane world cosmology. Then, in the case of left-handed
  sneutrino asymmetric dark matter, we find that the value of the lower
  bound on winos mass is smaller than that in the standard cosmological scenario.
 }
\end{center}
\end{titlepage}
\setcounter{footnote}{0}

\section{Introduction}

The cosmological and astrophysical data shows that the dark matter is the
dominant contribution of matter in the universe \cite{Planck:2018vyg}.
Unfortunately
the content of dark matter is not known until now. Particle physicists try
to find answer for this mystery beyond the standard model of particle
physics. Usually the neutral, long-lived or stable Weakly Interacting Massive
Particles (WIMPs) are considered to be excellent candidates for dark matter.
Among them, neutralino is the best motivated Majorana supersymmetric
particle for which the particle and anti-particle are the same. However, we
haven't detected the dark matter particles yet. Asymmetric Dark Matter (ADM) is
appeared as alternatives. If the asymmetry is
existed between the dark matter particle and anti-particle, it may
explain the similar order of the relic density of baryon and dark matter 
$\Omega_{\rm DM} \approx 4.7 \Omega_B$. There may exist common origin for this
coincidence. There are ADM models which aim to connect the
similarity between the dark matter and baryon asymmetry in the universe
\cite{adm-models}.

For ADM, the particles $\chi$ and anti-particles
$\bar\chi$ are distinct from each other. If they are annihilated sufficiently,
then the final relic density of ADM is determined by the asymmetry factor
which is the difference between the
particle and anti-particle comoving densities. Indeed the final relic density
is obtained by solving the Boltzmann equations for ADM.
In the standard
  framework of ADM relic density computation, it is assumed that ADM candidate
  particles WIMPs were
in thermal and chemical equilibrium state in the
radiation dominated epoch. When the interaction rate of ADM turns to
be smaller than the expansion rate of the universe, ADM candidate particles
decouple
from the equilibrium and their comoving number densities are almost fixed from
that freeze out point. Here 
the asymmetry is assumed to be existed before the decoupling of particles
and anti-particles from the equilibrium state and only the
annihilations from $\chi$ and $\bar\chi$ are considered
\cite{Graesser:2011wi,Iminniyaz:2011yp}. When the self annihilation is
permitted for ADM, there is
possibility that the pre-existed asymmetry between 
ADM particle and anti-particle could be washed out. 
Indeed some ADM models allow the self annihilation of ADM, for example, sneutino, higgsinos \cite{Kang:2011cni,Blum:2012nf}.
If the self annihilation rate of ADM particles
(anti-particles) at the temperature of the order of ADM mass is below the Hubble expansion rate, one can avoid the washing out
process. When the expansion rate of the universe becomes larger than the rate
of self annihilation, the wash-out processes freeze out.
Ref.\cite{Ellwanger:2012yg} derived the
upper bounds on self annihilation cross sections $\sigma_{\chi\chi} $
($\sigma_{\bar\chi\bar\chi}$) for ADM particle and
anti-particle  in the standard cosmology which is
used to limit the washing out of the pre-existed asymmetry.

On the other hand, we have no
  observational evidence of the radiation dominated era before the epoch of
  Big Bang Nucleosynthesis (BBN) where the temperature is above 1 Mev. There
  are many non-standard pre-BBN cosmological models which predicted the Hubble
  expansion rate is different from the standard one and it impacts the relic
  density of dark matter. Kination
is a period in which the energy density of universe is dominated by 
kinetic energy of the scalar field. The universe expands faster than the
standard cosmology in that era
\cite{Weinberg:2008zzc,Joyce:1996cp,Salati:2002md,DEramo:2017gpl,AristizabalSierra:2023bah,Co:2021lkc}. Faster
expansion rate leads to the earlier
decay of dark matter and resulted enhanced relic density
\cite{Salati:2002md,DEramo:2017gpl,Schelke:2006eg,Catena:2009tm}. In brane
world cosmology, standard model particles are assumed to be confined on 3-brane
when gravity lies in the whole higher dimensional spacetime \cite{Langlois:2002bb,Randall:1999vf,Binetruy:1999hy,Binetruy:1999ut}. Extra dimensions
impact the relic density of dark matter\cite{Okada:2004nc,AbouElDahab:2006glf}.
The relic density of ADM are affected by modification of the Hubble rate
\cite{Iminniyaz:2013cla,Wang:2015gua,Abdusattar:2015azp,Iminniyaz:2016iom,Iminniyaz:2018das}. In
previous work of the relic density calculation of ADM, it is assumed there is
annihilations only from ADM particle and anti-particle.
In our work, we investigate the relic abundance of ADM including self
annihilations in kination model
and brane world cosmology. We solve the coupled set of Boltzmann
equations which include the self annihilation cross sections for ADM particle $\chi$ and anti-particle $\bar\chi$ in non-standard
cosmological scenarios. The enhanced cosmic expansion rate causes the
freeze out point of wash-out to be earlier. We find the upper
bounds on self annihilation cross section, which allow the final asymmetry
to be sizable, is larger than that in the standard cosmology.

Our work is arranged as following, in the next section, we set up the
Boltzmann equations which include the self annihilations of $\chi\chi$ and
$\bar\chi \bar\chi$ in addition to the annihilations of $\chi\bar\chi$. In
section 3, the upper bounds on self annihilation cross section is obtained
in kination model and brane world cosmology. Then we discuss the
  constraints on winos mass in sneutrino ADM scenario based
  on the upper bounds of self annihilation cross section. The last section is devoted to the conclusions and discussions.

\section{Boltzmann equations for ADM including self
  annihilations  in standard cosmology}
The relic abundances of ADM particle and anti-particle are obtained by solving the
coupled set of Boltzmann equations for particle number density $n_{\chi}$ and
anti-particle number density $n_{\bar\chi}$ including self annihilations for
$\chi\chi$ and $\bar\chi\bar\chi$,
\begin{eqnarray} \label{eq:boltzmann_n}
  \frac{{\rm d}n_{\chi}}{{\rm d}t} + 3 H n_{\chi} &=&
       - \langle \sigma_{\chi\bar\chi} v\rangle
        (n_{\chi} n_{\bar\chi} - n^{\rm eq}_{\chi} n^{\rm eq}_{\bar\chi})\,
        - \langle \sigma_{\chi\chi} v\rangle
  (n^2_{\chi}  - n^{\rm eq \,2}_{\chi})\,;
  \nonumber \\
\frac{{\rm d}n_{\bar\chi}}{{\rm d}t} + 3 H n_{\bar\chi} &=&
   - \langle \sigma_{\chi\bar\chi} v\rangle (n_{\chi} n_{\bar\chi} - n^{\rm eq}_{\chi} n^{\rm eq}_{\bar\chi})\, - \langle \sigma_{\chi\chi} v\rangle
  (n^2_{\bar\chi}  - n^{\rm eq \,2}_{\bar\chi} )\,.
\end{eqnarray}
Here we assume $\sigma_{\chi\chi} = \sigma_{\bar\chi\bar\chi}$. The
equilibrium number densities are
\begin{eqnarray} \label{eq:boltzmann_neq}
n^{\rm eq}_{\chi} = g ~{\left( \frac{m T}{2 \pi} \right)}^{3/2}
  {\rm e}^{(-m + \mu)/T}\,,\,\,\,\,\,\,\,\,\,\,
  n^{\rm eq}_{\bar\chi} =  g ~{\left( \frac{m T}{2 \pi}
    \right)}^{3/2} {\rm e}^{(-m - \mu)/T}\,,
\end{eqnarray}
where $g$ denotes the number of internal degrees of freedom of the particle
$\chi$. We assume
$\chi$, $\bar\chi$ asymmetry is well created before the decoupling of 
ADM from the thermal equilibrium state. $H$
is the Hubble expansion rate and during the radiation dominated epoch
$H = \pi \sqrt{g_*/90} \, T^2/M_{\rm Pl} $  with
$M_{\rm Pl} = 2.4 \times 10^{18}$ being the reduced
Planck mass. $g_*$ is the effective number of relativistic degrees of
freedom.

We can rewrite the Boltzmann equations (\ref{eq:boltzmann_n}) in terms of the
dimensionless quantity $Y = n/s$ and $x = m/T$, with
$s = (2 \pi^2/45)\, g_{*s} T^3$ being the entropy density. Here $g_{*s}$ is
effective number of entropic degrees of freedom. Then
\begin{equation} \label{eq:boltzmann_Y}
  \frac{{\rm d} Y_{\chi}}{{\rm d}x} = -\frac{s}{Hx}\,
  \left[ 
      \langle \sigma_{\chi\bar\chi} v \rangle~
      (Y_{\chi}~ Y_{\bar\chi} - Y^{\rm eq}_{\chi}~Y^{\rm eq}_{\bar\chi}   )
      + \langle \sigma_{\chi\chi} v \rangle~
    (Y^2_{\chi} - Y^{\rm eq \,2}_{\chi}   ) \right]\, ;
\end{equation}
\begin{equation} \label{eq:boltzmann_Ybar}
\frac{{\rm d} Y_{\bar{\chi}}}{{\rm d}x}
= -\frac{s}{Hx}\,   \left[
  \langle \sigma_{\chi\bar\chi} v \rangle~
  (Y_{\chi}~Y_{\bar\chi} - Y^{\rm eq}_{\chi}~Y^{\rm eq}_{\bar\chi} )\,
   +  \langle \sigma_{\chi\chi} v \rangle~
  (Y^2_{\bar\chi} - Y^{\rm eq \,2}_{\bar\chi} )
  \right]\,.
\end{equation}
Here we assume $g_* \simeq g_{*s}$ and ${\rm d}{g_*}/{\rm d}x
\simeq 0$.
Usually, the thermal average of the annihilation cross section times relative
velocity of ADM is expanded as
\begin{equation} \label{eq:cross}
   \langle \sigma v \rangle = a + 6\,b x^{-1} + {\cal O}(x^{-2})\, .
\end{equation}
Here when $b = 0$, $s-$wave annihilation is dominant, and $a = 0$, 
$p-$wave annihilation is contributed.
\section{Relic abundance of ADM in  kination model  and
  brane world cosmology}
\subsection{Boltzmann equations in non-standard cosmological models}
The Hubble expansion rate in kination model is derived in
\cite{Salati:2002md} as
\begin{equation}\label{Hratio}
      H_k = H\,\sqrt{1 + \frac{\rho_{\phi}}{\rho_{\rm rad}}}\, ,
\end{equation}
where the ratio of scalar field energy density $\rho_{\phi}$ to the radiation
energy density $\rho_{\rm rad}$ can be written as 
\begin{equation}\label{rhoratio-1}
      \frac{\rho_{\phi}}{\rho_{\rm rad}} = \eta
      \left[ \frac{g_{*s}(T)}{g_{*s}(T_0)} \right]^2~
      \frac{g_*(T_0)}{g_*(T)}~
       \left( \frac{T}{T_0} \right)^2 \simeq \eta
       \left( \frac{T}{T_0}  \right)^2\, ,
\end{equation}
where the kination parameter $\eta = \rho_{\phi}(T_0)/\rho_{\rm rad}(T_0)$ \cite{Salati:2002md,Schelke:2006eg}. Here
the reference temperature $T_0$ is close to the freeze out temperature of WIMP annihilation and
\begin{equation}
      \rho_{\rm rad} = g_* (T) \frac{\pi^2}{30} T^4\,.
\end{equation}
Then
\begin{equation}\label{Hratio}
      H_k = H\,\sqrt{ 1+ \eta
       \left( \frac{x_0}{x}  \right)^2    }\,.
\end{equation}
Let us review the Hubble expansion rate in the brane world cosmology now.  
In the brane world cosmology, the Hubble expansion rate is expressed as
\begin{equation}
      H_{b} = H\, \sqrt{1 + \frac{k_b}{x^4}}\, ,
\end{equation} 
where $ k_b = \pi^2 g_* m^4 M^2_{\rm Pl}/(360 M^6_5) $, here $M_5$ is
the 5 dimensional Planck mass \cite{AbouElDahab:2006glf}.
The Boltzmann equations for the case of including self annihilations in 
 kination model and brane world cosmology become
\begin{equation} \label{eq:boltzmann_Yk}
  \frac{{\rm d} Y_{\chi}}{{\rm d}x} =
  -\frac{\lambda }{A_{k,b}}
  \left[ 
      \langle \sigma_{\chi\bar\chi} v \rangle
      (Y_{\chi} Y_{\bar\chi} - Y^{\rm eq}_{\chi}Y^{\rm eq}_{\bar\chi}   )
      + \langle \sigma_{\chi\chi} v \rangle
    (Y^2_{\chi} - Y^{\rm eq \,2}_{\chi}   ) \right]\, ;
\end{equation}
\begin{equation} \label{eq:boltzmann_Ykbar}
\frac{{\rm d} Y_{\bar{\chi}}}{{\rm d}x}
=  - \frac{\lambda }{A_{k,b}}
\left[
  \langle \sigma_{\chi\bar\chi} v \rangle
  (Y_{\chi}Y_{\bar\chi} - Y^{\rm eq}_{\chi}Y^{\rm eq}_{\bar\chi} )\,
   +  \langle \sigma_{\chi\chi} v \rangle
  (Y^2_{\bar\chi} - Y^{\rm eq \,2}_{\bar\chi} )
  \right]\,.
\end{equation}
Here $A_k = \sqrt{x^4 + \eta x_0^2x^2}$ and $A_b = \sqrt{x^4 + k_b}$,
$\lambda = 1.32\, m M_{\rm Pl}\sqrt{g_*} $ . Neglecting
${\cal O} \left((\mu/T)^2\right)$, $Y^{\rm eq}_{\chi}$ and $Y^{\rm eq}_{\bar\chi}$
become
\begin{equation}
     Y^{\rm eq}_{\chi} = 0.145\,\frac{g}{g_*}\, x^{3/2} {\rm e}^{-x} (1 +
     \frac{\mu}{T})\,;
\end{equation}
\begin{equation}
     Y^{\rm eq}_{\bar\chi} = 0.145\,\frac{g}{g_*}\, x^{3/2} {\rm e}^{-x} (1 
    - \frac{\mu}{T})\,.
\end{equation}
In the standard framework of ADM particle and anti-particle evolution, it is assumed that at
high temperature, ADM particles and anti-particles are in thermal
equilibrium in the early universe. When $T \lsim m$ and for
$m > |\mu|$, the equilibrium number densities of particle and
anti-particle drop exponentially. It causes the interaction rate
$\Gamma$ becomes smaller than the Hubble expansion rate $H$. Finally ADM
particles and anti-particles decouple from the equilibrium state and
their comoving number densities almost stay constant from the decoupling point
which is the freeze out temperature.

\subsection{The evolution of relic abundance of ADM}
For convenience, we express the Boltzmann equations (\ref{eq:boltzmann_Yk}) and 
(\ref{eq:boltzmann_Ykbar}) in terms of the difference and sum of $Y_{\chi}$
and $Y_{\bar\chi}$ as
\begin{equation}
  \Delta_- = Y_{\chi} - Y_{\bar\chi}\, ,\,\,\,\,\,
  \Delta_+ = Y_{\chi} + Y_{\bar\chi}\,.
\end{equation}
Indeed, $\Delta_-$ is the asymmetry factor which is related to the baryon
asymmetry. In \cite{Iminniyaz:2011yp}, the Boltzmann equations for particle
and anti-particle were solved by
numerically and analytically without self annihilation cross sections.
In that case,
$\Delta_-$ is constant. The final relic density of ADM is
determined by $\Delta_-$ and the annihilation cross section. When the self
annihilations are included in the Boltzmann equations, it is not
easy to solve Eqs. (\ref{eq:boltzmann_Yk}) and (\ref{eq:boltzmann_Ykbar}).

 For physical reasons, the self annihilation cross section can not be large, otherwise the asymmetry will be completely washed out.
  When $\sigma_{\chi\chi} \ll \sigma_{\chi\bar\chi}$, the term included
  $\sigma_{\chi\chi}$ can be neglected in the Boltzmann equations. In this case,
  adding
Eq.(\ref{eq:boltzmann_Yk}) to Eq.(\ref{eq:boltzmann_Ykbar}), we obtain 
\begin{equation}
  \frac{{\rm d} \Delta_+}{{\rm d}x} =
-\frac{1}{2}\,\frac{\lambda }{A_{k,b}}\,
  \langle \sigma_{\chi\bar\chi} v \rangle
  (\Delta^2_+- \Delta^2_- - {\Delta^{\rm eq}_+}^2 + {\Delta^{\rm eq}_-}^2)\,.
\end{equation}
here
\begin{equation}\label{eq:Delta+}
 \Delta^{\rm eq}_+ = 0.29\,\frac{g}{g_*} x^{3/2} {\rm e}^{-x}\,,\,\,\,\,\,
\Delta^{\rm eq}_- = 0.29\,\frac{g}{g_*} x^{3/2} {\rm e}^{-x}\,\frac{\mu}{T}.
\end{equation}
This is the case of only including ADM particle and anti-particle
annihilation $\sigma_{\chi\bar\chi}$. We will not
discuss it further here.

Subtracting Eq.(\ref{eq:boltzmann_Ykbar}) from Eq.(\ref{eq:boltzmann_Yk}), 
we obtain 
\begin{equation}\label{eq:Delta-}
  \frac{{\rm d} \Delta_-}{{\rm d}x} =
  -\frac{\lambda } {A_{k,b}}\,
  \langle \sigma_{\chi\chi} v \rangle
  (\Delta_+ \Delta_- - \Delta^{\rm eq}_+ \Delta^{\rm eq}_-),
\end{equation}

 For high temperatures $x < x_{fw}$, where $x_{fw}$ is the
  freeze out point for the self annihilation of ADM, $\Delta_+ $ follows its
  equilibrium value very closely, therefore, we can take the approximation
  $\Delta_+ \sim \Delta_+^{\rm eq}$, then Eq.(\ref{eq:Delta-}) becomes 
\begin{equation}\label{eq:Delta+simple}
  \frac{{\rm d} \Delta_-}{{\rm d} x} =
    - \frac{\lambda }{A_{k,b}}\,
    \langle \sigma_{\chi\chi} v \rangle \Delta^{\rm eq}_+
    (\Delta_- - \Delta^{\rm eq}_-).
\end{equation}  
For $T\sim m \,\, ( x \sim 1)$, the initial value of $\Delta_-$ is close to the equilibrium
value of $\Delta_-^{\rm eq}$ as $\Delta_{-{\rm in}} \sim \Delta_{ -}^{\rm eq} $.
Equation (\ref{eq:Delta+simple}) can be solved numerically with this initial
condition. Here we used Eq.(\ref{eq:cross}) and choose
$\mu/T=10^{-9}$, $x_{\rm in}=1$.


\begin{figure}[h] 
  \begin{center}
     \hspace*{-0.5cm} \includegraphics*[width=8cm]{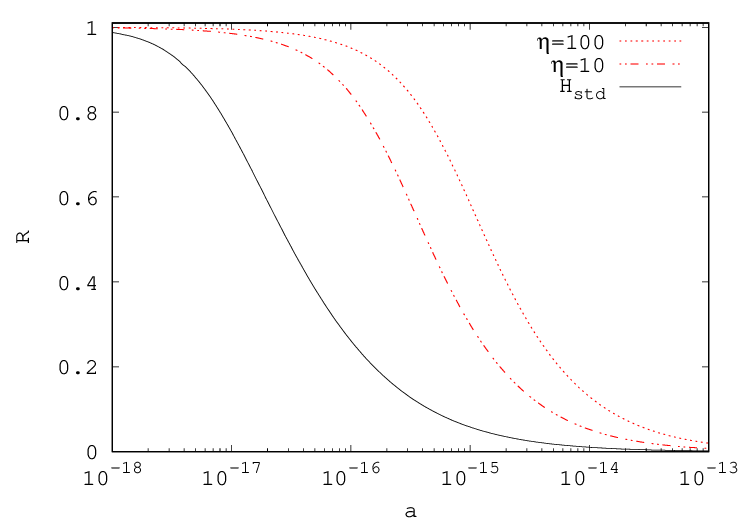}
    \put(-115,-12){(a)}
    \hspace*{-0.5cm} \includegraphics*[width=8cm]{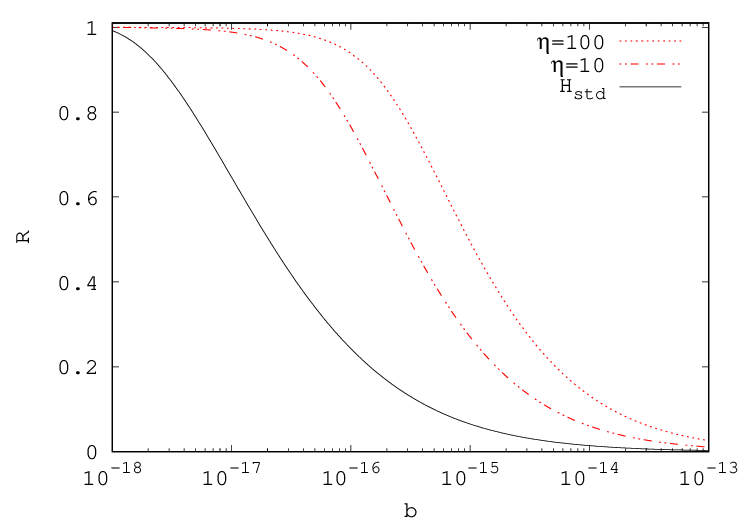}
    \put(-115,-12){(b)}
     \caption{\label{fig:a} \footnotesize
       The ratio of final asymmetry to the initial asymmetry
       $R = \Delta_{-\infty}/\Delta_{-\rm in}$ as a function of $a$
       or $b$ for the kination model and standard cosmology. Here $(a)$ is for
       $s-$wave annihilation and $(b)$ is for $p-$wave annihilation cross
       section. $m = 10$ GeV,
       $g = 2$, $g_* = 90$, $x_0=20$.  }  
   \end{center}
\end{figure}

In Fig.\ref{fig:a}, we plot the ratio of final asymmetry to the initial
asymmetry $R \equiv \Delta_{-{\rm \infty}}/\Delta_{-{\rm in}} $ as a function
of self annihilation cross section in kination model. For comparison, we
also plot this ratio for the standard cosmology.  
From  Fig.\ref{fig:a}, we find  the upper bounds on 
$\langle \sigma_{\chi\chi} v \rangle $ which allow the final
asymmetry to be sizable. When $R > 0.1$,
$a \lsim 4.4 \times 10^{-16}$ in the standard case. The upper bounds are
increased to $a \lsim 4.4 \times 10^{-15}$ for $\eta = 10$ and
$a \lsim 1.4 \times 10^{-14}$ for $\eta = 100$. In other words,
when the ratio of final asymmetry to the initial asymmetry is larger than
0.1, there is still asymmetry exists and it is not completely washed out by
the self annihilation. For $p-$wave
annihilation cross section, the limits are slightly different from 
$s-$wave annihilation, for example, when $R > 0.1$,  $b \lsim 4.8 \times 10^{-16}$ in the
standard cosmology and $b \lsim 4.6 \times 10^{-15}$ for $\eta = 10$ and
$b \lsim 1.5 \times 10^{-14}$ for $\eta = 100$. In kination model, when the
modified factor $\eta$ is larger, the upper
bound on the self annihilation cross section is larger. 

\begin{figure}[h]
  \begin{center}
     \hspace*{-0.5cm} \includegraphics*[width=8cm]{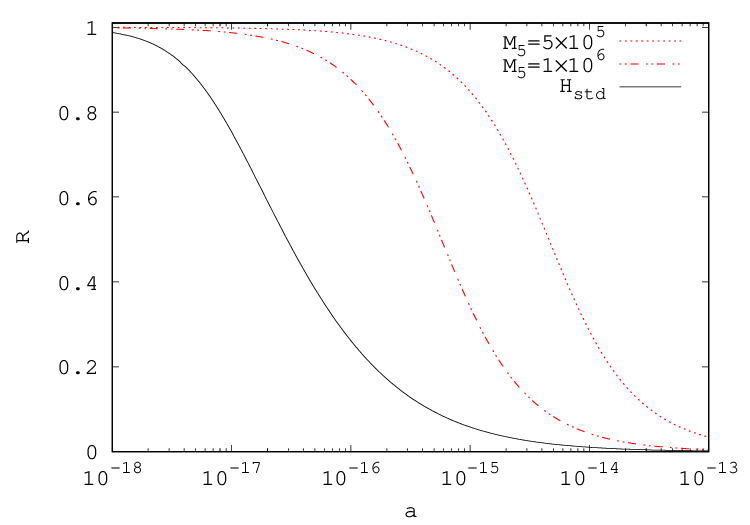}
    \put(-115,-12){(a)}
    \hspace*{-0.5cm} \includegraphics*[width=8cm]{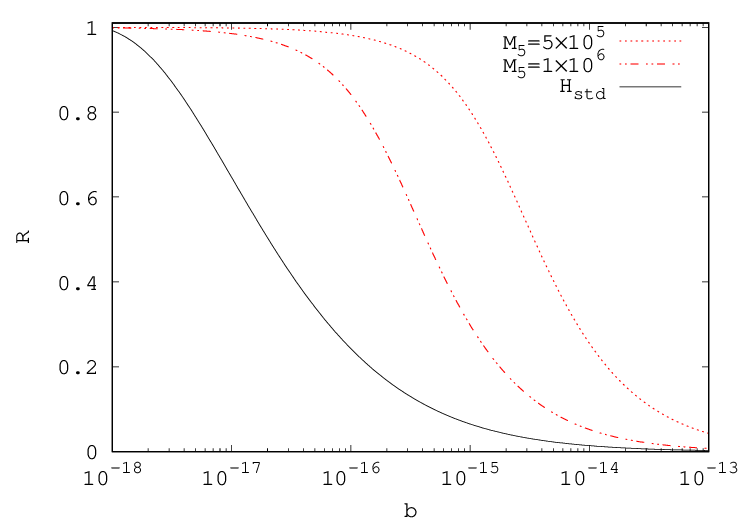}
    \put(-115,-12){(b)}
     \caption{\label{fig:b} \footnotesize
       The ratio 
       $R $ as a function of $a$ or $b$ for the brane world cosmology and
       standard cosmology. Here $(a)$ is
       for $s-$wave annihilation and $(b)$
       is for $p-$wave annihilation cross section. 
       $m = 10$ GeV, $g = 2$, 
  $g_* = 90$.  }  
   \end{center}
\end{figure}

The same rule is applied to the case of brane world cosmology.
The ratio $R$ as a function of self annihilation cross section in brane
world cosmology is shown in Fig.\ref{fig:b}. While $R > 0.1$, the upper bounds
are $a \lsim 4.0 \times 10^{-15}$ for $M_5 = 10^6$ and
$a \lsim 3.2 \times 10^{-14}$ for $M_5 = 5 \times 10^5$ in the case of
$s-$wave annihilation;
for $p-$wave annihilation, $b \lsim 4.4 \times 10^{-15}$ for $M_5 = 10^6$ and
$b \lsim 3.5 \times 10^{-14}$ for $M_5 = 5 \times 10^5$. In brane world
cosmology, when $M_5$ is smaller, the upper bound on self annihilation cross
section is larger.

Following the assumption that at high temperature, ADM particles and
anti-particles are approximately in the thermal equilibrium state,
and $Y_{\chi(\bar\chi)}$ is
close to $Y_{\chi(\bar\chi)}^{\rm eq}$, then
\begin{equation}
  (\Delta_--\Delta^{\rm eq}_-)^2\approx 0 \, ,
\end{equation}
therefore, Eq.\eqref{eq:Delta+simple} can be written as
\begin{equation}\label{delta2}
	\frac{{\rm d} \Delta_-}{{\rm d} x} =
	- \frac{\lambda }{2A_{k,b}\,\mu/T}\,
	\langle \sigma_{\chi\chi} v \rangle 
	({\Delta_-}^2 - {\Delta^{\rm eq}_-}^2).
\end{equation}
Eq.\eqref{delta2} has the same mathematical form as the known Boltzmann
equation which describes the symmetric dark matter particle abundance evolution
\cite{Scherrer:1985zt}. This
situation lets us calculate the scaled freeze out
temperature $x_{fw}$ of wash-out processes by following the standard picture
of the dark matter particle evolution. It is assumed the freeze out occured
when the deviation $\delta\equiv\Delta_--\Delta^{\rm eq}_-$ is of the same
order as the equilibrium value of $Y_{\chi,\bar\chi}$ ,
\begin{equation}\label{denxf}
\delta(x_{fw})=\xi\Delta^{\rm eq}_-(x_{fw}) \:,
\end{equation}
where $\xi=\sqrt{2}-1$ \cite{Scherrer:1985zt}.

 Eq.\eqref{eq:Delta+simple} is written as 
\begin{equation}\label{Ddelta}
\frac{\rm d \delta}{{\rm d}x}=-\frac{\rm d\Delta^{\rm eq}_-}{{\rm
    d}x}-\frac{\lambda\langle
  \sigma_{\chi\chi}v\rangle}{A_{k,b}}\frac{T}{\mu}\Delta^{\rm eq}_- \, \delta \,.
\end{equation}
The approximate solution for Eq.\eqref{Ddelta} is
\begin{equation}\label{delta}
\delta=\frac{ A_{k,b}\mu}{\lambda  \langle \sigma_{\chi\chi} v\rangle T} \ ,
\end{equation}
where we used ${\rm d \delta}/{\rm d}x\approx0$ and ${\rm d\Delta^{\rm eq}_-}/{\rm d}x\approx -\Delta^{\rm eq}_-$.
Substituting Eq.\eqref{delta} into Eq.\eqref{denxf}, 
we find that the freeze out point $x_{fw}$
 of wash-out processes has the same
form as the standard freeze out temperature of WIMP annihilation,
\begin{equation}\label{xf}
 x_{fw}=\ln\frac{0.29\,\xi\lambda g\langle\sigma_{\chi\chi}v\rangle}{g_*x^{1/2}(A_{k,b}/x^2)} \  \Bigg|_{x=x_{fw}}\,.
\end{equation}

We can compare $R$ which is obtained by the numerical solutions  of
Eq.\eqref{eq:Delta+simple} with the
approximate solutions $R_{\rm app}(x_{fw})$,
\begin{equation}\label{Rapprox}
 R_{\rm app}(x_{fw}) \equiv \frac{\Delta^{\rm eq}_-(x_{fw})}{\Delta^{\rm eq}_-(1)}\approx \frac{x_{fw}^{3/2}e^{-x_{fw}}}{e^{-1}}\,.
\end{equation}
The freeze out points of wash-out provided by Eq.\eqref{xf} are
$x_{fw} = 5.9,\,\, 5.8, \,\, 5.7$ respectively for $a=4.4\times
10^{-16}$ in the standard cosmology, $a=4.4\times 10^{-15}$
in kination model with $\eta=10$, and $a=4.0\times 10^{-15}$ in the
brane world cosmology with $M_5=10^6$ when $R=0.1$. Substituting $x_{fw}$
into Eq.\eqref{Rapprox}, the results are
$R_{\rm app} = 0.11,\,\, 0.11,\,\, 0.12$. We find the approximate
solutions for $R$ match with the numerical results well.

\begin{figure}[h]
  \begin{center}
     \hspace*{-0.5cm} \includegraphics*[width=8cm]{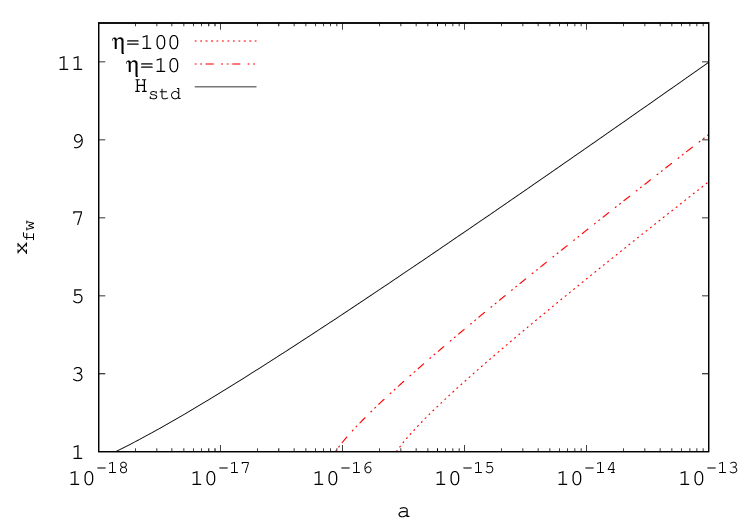}
    \put(-115,-12){(a)}
    \hspace*{-0.5cm} \includegraphics*[width=8cm]{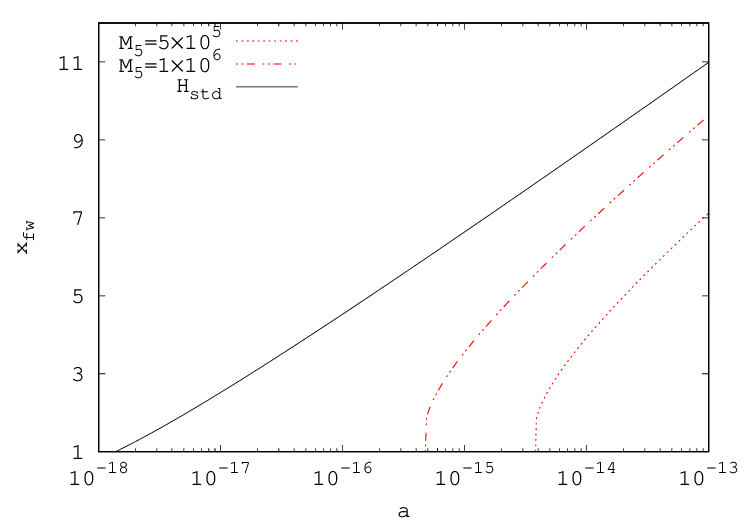}
    \put(-115,-12){(b)}
     \caption{\label{fig:c} \footnotesize
       The scaled freeze out temperature $x_{fw}$ of wash-out processes as a
       function of self annihilation cross section. Here, $(a)$ is for the
       $s-$wave self annihilation cross section in the kination model 
       and $(b)$ is in the brane world cosmology. $m = 10$ GeV, $g = 2$,
       $g_* = 90$, $x_0=20$.}  
   \end{center}
\end{figure}

The scaled freeze out temperature $x_{fw}$ of wash-out processes as a function
of self annihilation cross section $a$ is plotted in Fig.\ref{fig:c}.
It shows that, for a certain value of $x_{fw}$, models with enhanced
expansion rate allow the larger self annihilation cross sections to
exist.  For the greater levels of modification of expansion rate,
  there is larger bound for self annihilation cross section. If we look at the
following equation
\begin{equation}\label{xfwashout}
	H(T_{fw})=(n_\chi^{\rm eq}-n_{\bar{\chi}}^{\rm eq})\langle \sigma_{\chi\chi} v\rangle (T_{fw}) \ ,
\end{equation}
we find for a certain value of $x_{fw}$, expansion rate $H$ in the left
hand side of Eq.\eqref{xfwashout} is enhanced in kination model and brane
world cosmology respectively. The equivalence requires that
$\sigma_{\chi\chi}$ in the right hand
side of Eq.\eqref{xfwashout} to be increased. In other words, the
wash-out processes in kination model and brane world cosmology freeze out
earlier than that in the standard cosmology, it resulted the larger
final asymmetry of ADM. Therefore, the enhanced cosmic expansion rate allows
the larger self annihilation cross section to exist.

\subsection{The constraints on the winos mass  in sneutrino ADM}

 In this subsection, we study the effects of upper bounds on
 $\sigma_{\chi\chi}$ for sneutrino ADM. 
The left-handed sneutrino is assumed to be ADM, which has been proposed in refs.\cite{Kang:2011cni,Hooper:2004dc,Ma:2011zm}.
In sneutrino ADM scenarios, by exchanging of the electroweak
gauginos in t-channel, the left-handed sneutrinos $\tilde{\nu}_L$ can self
annihilate through the processes of
$\tilde{\nu}_L + \tilde{\nu}_L \rightarrow \nu_L + \nu_L $. 
 Electroweak gauginos are the binos with mass $M_1$ and winos with mass $M_2$.
 For $M_1 , M_2 \gg m $, the corresponding expression for $\langle\sigma_{\tilde{\nu}_L \tilde{\nu}_L} v\rangle$ is \cite{Nihei:2002sc}
\begin{equation}
\langle\sigma_{\tilde{\nu}_L \tilde{\nu}_L} v\rangle \simeq \frac{g_2^4}{16\pi}(1-\frac{3}{2x})(\frac{\tan^2 {\theta_W}}{M_1}+\frac{1}{M_2})^2\ ,
\end{equation} 
where $g_2=e/\sin\theta_W$ is $SU(2)$ gauge coupling and $\theta_W$ is the weak mixing
angle, $e$ is the charge of electron. With the assumptions that the universal gaugino masses at the GUT
scale, $M_1$, $M_2$ are correlated by $M_1\simeq M_2/2$ and with
$\tan^2\theta_W\approx0.3$, the expression for $\langle\sigma_{\tilde{\nu}_L
  \tilde{\nu}_L} v\rangle$ can be written as \cite{Ellwanger:2012yg}
\begin{equation}\label{winocross}
\langle\sigma_{\tilde{\nu}_L \tilde{\nu}_L} v\rangle \simeq \frac{g_2^4}{8\pi}(1-\frac{3}{2x})\frac{1}{M_2^2} \ .
\end{equation}
In the previous subsection, we deduced the upper bounds on
  self annihilation cross section $\sigma_{\chi\chi}$ in kination and brane
  world cosmology. Here we use the limitations for $\sigma_{\chi\chi}$ to
find constraints on the winos mass.
From Fig.\ref{fig:a} and Fig.\ref{fig:b}, we can see the first term in
Eq.\eqref{winocross} gives rise
to the  stronger constraints and applying the bound R$>$0.1 in kination and
brane world cosmological scenarios, we obtain
\begin{equation}
\begin{split}
	&M_2\gsim1\times10^7 {\rm Gev} \times (\frac{m}{100\ {\rm Gev}})^{1/2} ,
\qquad  \:\eta=10, \\
&M_2\gsim 7\times10^6 {\rm Gev} \times (\frac{m}{100\ {\rm Gev}})^{1/2},
\qquad  \:\eta=100 ,\\
&M_2\gsim1\times10^7 {\rm Gev} \times (\frac{m}{100\ {\rm Gev}})^{1/2},
\qquad \:M_5=10^6, \\
&M_2\gsim4\times10^6 {\rm Gev} \times (\frac{m}{100\ {\rm Gev}})^{1/2} ,
\qquad  \:M_5=5\times10^5. \\
\end{split}
\end{equation}
In the standard cosmological
scenario, ref.\cite{Ellwanger:2012yg} obtained the mass range for $M_2$ as
\begin{equation}
M_2\gsim3\times10^7 {\rm Gev} \times (\frac{m}{100\ {\rm Gev}})^{1/2}\,\,\,\,\,
({\rm standard\  case}).
\end{equation}
We find when the cosmic expansion rate is increased in kination and brane world
cosmology, the value for the lower bound on $M_2$ is smaller than that in the
standard cosmological scenario. Because of the enhanced expansion rate in
kination and brane world cosmology, 
ADM particles and anti-particles decouple earlier than that in the standard
cosmology. It leads to the larger
upper bound for self annihilation cross section. The larger upper bounds on
  $\sigma_{\chi\chi}$ result in the smaller values of lower bounds on
  winos mass.
\section{Summary and conclusions}

We investigated the relic abundance of ADM including
self annihilation in kination and brane world cosmological scenarios. If the
self annihilation is allowed, there is possibility that the pre-existed
asymmetry will be washed out. We found when the ratio $R$ of final
asymmetry $\Delta_{-\infty}$ to the initial asymmetry $\Delta_{-\rm in}$ is
sizable, the upper bounds on self annihilation cross section in
kination and brane world cosmological models are larger than that in the
standard cosmology. Applied the constraints on self annihilation
 cross section to the sneutrino ADM, we found that the larger upper bounds on
  $\sigma_{\chi\chi}$ lead to the smaller values of lower bounds on
  winos mass.
We also calculated the effect of enhanced cosmic expansion
rate on the freeze out point of wash-out. The enhanced cosmic expansion rate
causes the
freeze out of wash-out process to be earlier, then the larger final
asymmetry is subsequently preserved.  Therefore, the larger self
annihilation cross section is allowed to exist in the non-standard
cosmological scenarios. 

Our result is important to know what extent the self annihilation cross
sections should be in order to the final asymmetry be sizable in the
non-standard cosmological scenarios including kination and brane world
cosmological models. The conclusion is helpful for ADM models which neglected
the self annihilation processes in the past.

\section*{Acknowledgments}

The work is supported by the National Natural Science Foundation of China
(2020640017, 11765021, 2022D01C52).

\end{document}